\documentclass[epj]{svjour}

\usepackage{graphicx}% Include figure files
\usepackage{bm}% bold math

\usepackage{color}
\newenvironment{ckomm}{}{ }
\newcommand{\QQ}{\begin{ckomm}\color{red} \bf}
\newcommand{\QQEND}{\end{ckomm}}
\begin{document}
\title{Loading of a cold atomic beam into a magnetic guide}
\titlerunning{Loading of an atomic beam into a magnetic guide}

\author{P. Cren, C. F. Roos, A. Aclan, J. Dalibard and D. Gu\'ery-Odelin}
\institute{Laboratoire Kastler Brossel\thanks{Unit\'e de Recherche
de l'Ecole Normale Sup\'erieure et de l'Universit\'e Pierre et
Marie Curie, associ\'ee au CNRS.
}, D\'epartement de Physique de l'Ecole Normale Sup\'{e}rieure, \\
24 rue Lhomond, 75005 Paris, France}

\date{\today}

\abstract{ We demonstrate experimentally the continuous and pulsed loading of a slow and cold atomic
beam into a magnetic guide. The slow beam is produced using a vapor loaded laser trap, which ensures
two-dimensional magneto-optical trapping, as well as cooling by a moving molasses along the third
direction. It provides a continuous flux larger than $10^9$ atoms/s with an adjustable mean velocity
ranging from 0.3 to 3 m/s, and with longitudinal and transverse temperatures smaller than
$100\;\mu$K. Up to $3\; 10^8$ atoms/s are injected into the magnetic guide and subsequently guided
over a distance of 40 cm.}


\maketitle

\section{Introduction}
\label{sec:introduction}

A large variety of techniques based on the use of electromagnetic forces are now available to control
the motion of atoms \cite{Chu98,Cohen98,Phillips98} and molecules
\cite{Fioretti,Takekoshi98,Weinstein98,Bethlem00}. The progress in this domain has opened the way to
several fields of applications such as atom optics and interferometry \cite{Cargese00}, precision
experiments and metrology \cite{Santarelli99}, and statistical physics, with the achievement of
degenerate Bose or Fermi gases \cite{Varenna98}. Among these techniques, a special effort has been
devoted to the production of slow and cold continuous atomic and molecular beams, since they
represent a powerful tool for many of the above mentioned applications.

A spectacular challenge in this field consists in the achievement
of a continuous beam operating in the quantum degenerate regime.
This would be the matter wave equivalent of a cw monochromatic
laser and it would allow for unprecedented performances in terms
of focalization or collimation. A possible way to achieve this
goal has been studied theoretically in \cite{Mandonnet00}. In this
proposal, a non-degenerate, but already slow and cold beam of
particles, is injected into a magnetic guide
\cite{Schmiedmayer95,Denschlag99,Goepfert99,Key00,Dekker00,Teo01,Sauer01,Hinds99}
where transverse evaporation takes place. If the elastic collision
rate is large enough, efficient evaporative cooling will lead to
quantum degeneracy after a propagation length in the guide
compatible with experimental constraints (\textit{i.e.} a few
meters). Therefore the success of this project relies upon two
preliminary and separate accomplishments. First, one has to build
a source of cold atoms as intense as possible, and secondly, one
has then to inject the atoms into a magnetic guide, where they
should propagate with reduced losses over a long distance.

When resonant laser light is available at the atomic resonance
frequency, a convenient method for producing a bright source of
slow atoms consists in decelerating a thermal beam with radiation
pressure \cite{Prodan82,Ertmer85}, and subsequently applying
transverse cooling and trapping to the slow atomic beam (``atom
funnel") \cite{Riis90,Nelessen90,Swanson96,Lison99,Chen00}.
Alternatively, one can extract a jet of atoms from a vapor-loaded
magneto-optical trap (MOT). The simplest scheme consists in a 2D
MOT \cite{Dieckmann98,Pfau02} where the MOT laser beams propagate
only in the $xy$ plane, resulting in a uncooled atom jet along the
$z$ axis. As for some of the demonstrated funnels
\cite{Riis90,Swanson96,Chen00}, one can narrow the longitudinal
velocity distribution, using a moving molasses scheme,
\emph{i.e.}, a pair of counter-propagating laser beams along the
atomic beam axis, with two different frequencies \cite{Weyers97}.
The atoms are then bunched in a non-zero velocity class, which
corresponds to the moving frame where both lasers are seen with
equal frequencies. One can also produce a beam with two or more
velocity components using a 2D ($xy$) magneto-optical trap and a
static magnetic field superimposed with a static optical molasses
along the beam axis ($z$) \cite{Weyers97,Dudle96,Berthoud98}.
Alternatively a collimated beam of atoms can be obtained by
drilling a hole in one of the mirrors used for the MOT \cite{Lu96}
(see also \cite{Dieckmann98}), with a pyramidal mirror structure
with a hole at its vertex \cite{Lee96,Williamson98,Arlt98}, or
with an extra pushing beam, which destabilizes the MOT at its
center \cite{Wohlleben01,Cacciapuoti01}. All these sources produce
a relatively bright beam (from 10$^6$ to several $10^{10}$ atoms
per second), with an average velocity $\bar v$ between 1 and
50~m/s; for most of them the velocity distribution is rather
broad, with a dispersion $\Delta v\sim \bar v$.

In the perspective of loading a magnetic guide in order to perform evaporative cooling, the atom
source should provide an average velocity $\bar v$ smaller than 2~m/s, and it should be as cold as
possible, both longitudinally and transversally ($\Delta v \ll \bar v$). This eliminates most of the
source types described above, in particular those from the ``leaking MOT" family, and also the
sources producing a beam with multiple velocity components. One is then left with the principle of a
source which uses a moving molasses (MM) on axis, together with a transverse magneto-optical trap
(MOT), which we shall designate in the following as a ``MM-MOT".

In the present paper, we describe in \S~\ref{sec:MMMOT} our vapor loaded MM-MOT, which produces a
very slow and cold atomic beam ($\bar v$ between 0.3 and 3 m/s, $\Delta v < \bar v/10$), and which is
well suited for loading  a magnetic guide, either continuously or in a pulsed manner. We also compare
our results with previously reported work. In \S~\ref{sec:guide}, we show how to inject this atomic
beam into a magnetic guide, created by a quadrupole field in the $xy$ plane. We discuss in the
concluding section (\S~\ref{sec:conclusion}) the perspectives opened by this experimental setup for
the development of a continuous, evaporatively cooled, atomic beam. The present performances of our
system are still far from the requirements for initiating forced evaporative cooling in the magnetic
guide. However, with an improved setup,we hope to reach the desired conditions as discussed in the
concluding section.

\section{The ``MM - MOT"}
\label{sec:MMMOT}

In this section we describe the principle and the practical realization of the MM-MOT, whose purpose
is to provide a cooling along the $z$ axis, in the moving frame defined by the moving molasses (MM),
and a magneto-optical trapping (MOT) in the $xy$ plane. This setup is employed to capture $^{87}$Rb
atoms from the background gas.
\subsection{Principle of the MM-MOT}
\label{subsec:PrincipleMMMOT}

The MM-MOT  is based upon a four-beam laser configuration similar to the one used for the study of
optical lattices described in \cite{Grynberg93}, superimposed with a linear magnetic 2D quadrupole
field.

The magnetic quadrupole field is created by four rectangular, elongated coils located in the planes
$x=\pm 8$~cm and $y=\pm 8$~cm. The currents in the coils facing each other have opposite sign so
that, close to the symmetry axis $x=y=0$, the resulting field is given by \[ {\bf B}=(-b'x\; ,\; b'y\
\;, \;0)\ .
\]
The field is zero along the $z$ axis and the transverse gradient
is typically $b'=0.1$~T/m.

\begin{figure}
 \includegraphics[width=0.45\textwidth]{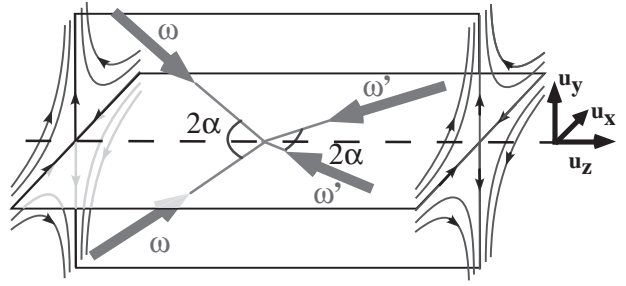}
 \caption{Laser configuration of the 2D MOT}
 \label{fig:configuration}
\end{figure}

The optical arrangement consists of four laser beams in a
tetrahedral configuration (see Fig. \ref{fig:configuration}). Two
laser beams with frequency $\omega$ propagate in the $yz$ plane
along the directions $\cos\alpha \;{\bf u}_z \pm \sin \alpha\;
{\bf u}_y$, with a positive helicity. The two other beams with
frequency $\omega'$ propagate in the $xz$ plane along the
directions $-\cos\alpha\; {\bf u}_z \pm \sin \alpha \;{\bf u}_x$,
with a negative helicity. The four beams are red-detuned with
respect to the atomic transition\\
$|5S_{1/2},F=2\rangle\rightarrow |5P_{3/2},F=3\rangle$, whose frequency is called $\omega_a$. The
average detuning $\delta=\bar \omega -\omega_a$, with $\bar \omega=(\omega+\omega')/2$, is typically
$-3\;\Gamma$, where $\Gamma=2\pi\times 5.9$~MHz denotes the natural width of the excited level of the
transition.

For $\alpha=\pi/2$ and $\omega=\omega'$, one obtains the ``standard" 2D-MOT configuration
\cite{Dieckmann98}, that confines and cools the atoms transversally without affecting the
longitudinal motion. For $\alpha \sim \pi/4$, one obtains an efficient cooling of the atomic motion
along the $z$ direction, while still keeping an efficient magneto-optical trapping in the $xy$ plane.
In our experiment we work with $\alpha=47\,^\circ$.

Because of the frequency difference $\omega-\omega'$ between the
two pairs of beams, the cooling corresponds to an accumulation of
the atoms around the velocity class
\begin{equation}
\bar{v}=\frac{\omega-\omega'}{2k\cos\alpha} \ ,
 \label{eq1}
\end{equation}
where $k$ is the wave vector of the laser light. By scanning $\omega-\omega'$ between 0 and $2\pi
\times 5$~MHz (\emph{i.e.} a small fraction of the detuning $\delta$), one can adjust the mean
velocity $\bar v$ between 0 and $\sim 3$~m/s.

We have confirmed this qualitative understanding of the MM-MOT using a numerical integration of the
trajectories of the atoms in the corresponding laser configuration. This treatment follows the same
lines as in \cite{Wohlleben01}. We model the atomic transition as a $|g,J_g=0\rangle \leftrightarrow
|e,J_e=1\rangle $ transition with frequency $\hbar \omega _a$, where $|g\rangle $ and $|e\rangle $
stand for the ground and excited state respectively. We denote by $I_{\rm sat}=16.2$~Wm$^{-2}$ the
saturation intensity for this transition. For each of the laser beams ($j=1,\ldots,4$) and for each
of the transitions $|g\rangle \leftrightarrow |e_m\rangle$ ($m=-1,0,1$), we introduce the position
and velocity dependent saturation parameter
\[
s_{j,m}=\frac{I_{j,m}}{I_{\rm sat}}\;\frac{\Gamma ^2}{\Gamma ^2 +
4(\delta _j-\mathbf{k}_j\cdot \mathbf{v}+ m\mu B/\hbar )^2}\ .
\]
Here $\mu$ is the magnetic moment associated with the level $|e\rangle$, $B$ is the local magnetic
field amplitude, and ${\bf k}_j$, $\delta_j$ and $I_{j,m}$ denote the wave vector, detuning and local
intensity of the $j$-th laser wave, driving the $|g\rangle \leftrightarrow |e_m\rangle$ transition.
We then choose the following approximate expression for the total radiative force acting on an atom:
\begin{equation}
{\bf F}=\sum_i \hbar {\bf k}_i~\frac{\Gamma}{2}~\frac{\sum_m
s_{i,m}}{1+\sum_{j,m} s_{j,m}}\ .
 \label{Fsum}
\end{equation}
A detailed discussion of the approximations leading to (\ref{Fsum}) can be found in
\cite{Wohlleben01}. Here we simply recall that this expression corresponds to the spatial average of
the radiative force over a cell of size $\lambda^3 =(2\pi /k)^3$, so that all interference terms
varying as $ê{i(\mathbf{k}_j-\mathbf{k}_{j^{\prime }})\cdot \mathbf{r}}$ have been neglected.

In the simulation the initial position of each atom is chosen following a uniform spatial
distribution on the walls of the vacuum vessel . The initial velocity is given by the Maxwell
Boltzmann distribution for $T=300$\,K. By computing a large number of trajectories, one obtains the
probability for an atom to be captured and transferred into the jet, as well as the jet
characteristics which are the velocity distribution, the divergence, and the total flux. The total
flux of the simulated jet is calculated using the real number of atoms $\dot{\cal N}$ emitted per
unit time and per unit surface of the cell at a pressure $P$. One obtains $\dot{\cal N}= P/\sqrt{2\pi
m k_{\rm B} T}$ \cite{Reif}. Note that the simulation neglects interaction effects like collisions
and multiple light scattering. The validity of the linear scaling with pressure is limited to the low
pressure regime ($P<10^{-7}$\,mbar) where the characteristic time of $\sim 20$\,ms for extracting an
atom out of the MM-MOT is shorter than the collision time.

In addition to confirming the qualitative understanding of the MM-MOT, this simulation is
particularly helpful for optimizing certain parameters for which a fine tuning on the real
experimental setup would be tiresome. In particular we used it for the determination of the optimal
laser waists for the experimentally available laser power.

\begin{figure}
\includegraphics[width=0.45\textwidth]{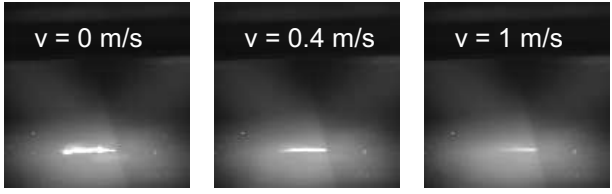}
\caption{Photograph of the 2D MOT for various mean velocities. From left to right: $\bar v=0, 40,
100$~cm/s.}
 \label{fig:photo}
\end{figure}

\subsection{Experimental realization of the MM-MOT}

The trap is created in a rectangular glass cell (130~mm$\times$ 50~mm $\times$ 50~mm) and the atoms
are captured from the low-pressure background gas. The partial pressure for $^{87}$Rb is measured by
the absorption of a resonant beam. It can be varied from $10^{-9}$ mbar up to the saturated vapor
pressure at room temperature ($3\times 10^{-7}$ mbar) by controlling the temperature of the
Rb-reservoir or the aperture of the intermediate valve.

The laser light for the magneto-optical trap originates from four laser diodes (one per beam), which
are injection-locked to a master diode. Each trapping beam has a circular beam profile with a waist
of about 15 mm and a power of 30 mW at the level of the trap. The linewidth of the master diode is
narrowed by optical feedback from an external grating. It is split into two beams, each of which
passes through an acousto-optic modulator, which provides the frequencies $\omega$ and $\omega'$. The
stability of $\omega-\omega'$ is better than 10~kHz, which defines the velocity $\bar v$ to better
than 0.5~cm/s.

A second grating stabilized laser diode, resonant with
$|5S_{1/2},F=1\rangle\rightarrow |5P_{3/2},F=2\rangle$,
illuminates the MM-MOT. This repumping laser prevents the
accumulation of atoms in the ground level $F=1$ by optical
pumping. Both master lasers are frequency stabilized using
saturated absorption spectroscopy in a rubidium vapor cell.

Photographs of the fluorescence light emitted by the trapped atoms
are shown in Fig.~\ref{fig:photo} for various mean velocities
$\bar v$. We observe elongated clouds of atoms with a length
$L\sim 20$~mm and a transverse size $\sim 500\;\mu$m. Due to local
imperfections of the intensity profile of the laser beams, the
atomic cloud does not have a straight shape over its whole length
when it is at rest ($\omega=\omega'$). One rather observes a
wiggling line with an amplitude $\sim 500\ \mu$m
(Fig.~\ref{fig:photo}a). These wiggles disappear as soon as one
imposes a non-zero mean velocity $\bar v$ on the atoms
(Fig.~\ref{fig:photo}b,c).

\subsection{Characterization of the MM-MOT}

The characterization of this trap consists in determining the main
features of the atomic beam that is produced: mean velocity,
longitudinal and transverse temperatures, and flux. The first
quantities are determined from the absorption of a probe beam
located at a distance $z_0$ away from the output of the MM-MOT,
while the latter is inferred from the fluorescence light emitted
by the atoms in the trap region.

\subsubsection{Longitudinal velocity distribution}

In order to investigate the velocity distribution of the atomic beam along the $z$ axis, we use a
time-of-flight technique. We operate the MM-MOT with a plug laser beam present which blocks the
outgoing atomic beam by pushing the atoms away from the $z$ axis. At time $t=0$, we switch off the
plug beam, so that the atomic beam can propagate freely and we monitor as a function of time the
modification of the absorption of a probe beam located further downstream.

The probe beam is tuned to the $|5S_{1/2},F=2\rangle\rightarrow |5P_{3/2},F=3\rangle$ transition,
 superimposed with a weak repumping laser to overcome optical pumping effects. The waist of the probe
beam is 160 $\mu$m and its distance $z_0$ from the end of the MM-MOT is 8.3~mm. This distance is
large in comparison with the diameters of the plug and probe beams and has been chosen in order to
overcome spatial size effects.

\begin{figure}
 \includegraphics[width=0.45\textwidth]{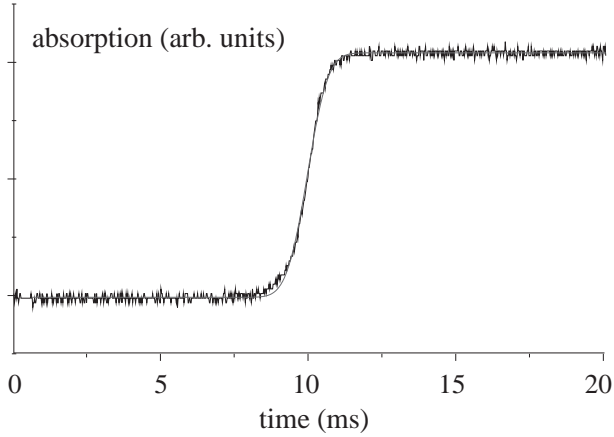}
\caption{Time of flight signal from the MM-MOT. The straight line
is a fit using (\ref{eq2}) with $\bar{v}=84\;$cm/s and $\Delta
v_\parallel=4.6\;$cm/s.}
 \label{fig:tof}
\end{figure}

The absorption of the probe beam is detected by a photodiode. A typical step-shaped signal is given
in Fig.~\ref{fig:tof}. The solid curve is a fit from the relation
\begin{equation}
s(z_0,t) \propto  1-\mbox{Erf}
\bigg(\frac{z_0-\bar{v}t}{\sqrt{2}\Delta v_{\parallel} t}\bigg) \
,
 \label{eq2}
\end{equation}
which corresponds to the expected time-of-flight signal assuming a
Gaussian velocity distribution centered on $\bar v$ and with a
dispersion $\Delta v_\parallel$. We find generally an excellent
agreement between this fitting function and the experimental
signal. For the particular case of Fig.~\ref{fig:tof}, we obtain
$\bar v=84\;$cm/s and $\Delta v_\parallel=4.6\;$cm/s,
corresponding to a longitudinal temperature $T_{\parallel}=m\;
\Delta v_\parallel^2/k_B =23\;\mu$K.

\begin{figure}
\includegraphics[width=0.45\textwidth]{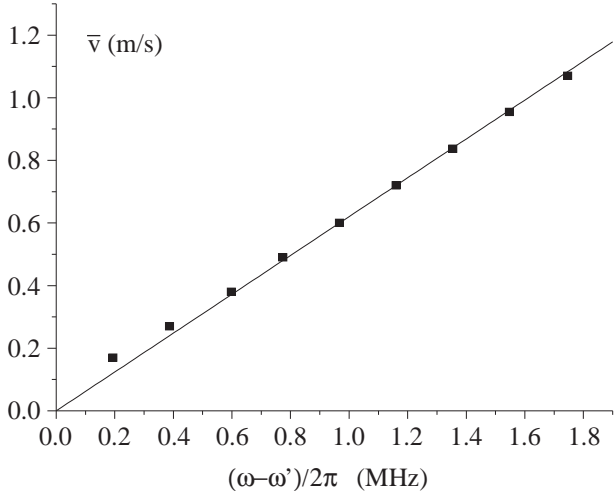}
\caption{Mean velocity $\bar v$ of the beam as a function of the
frequency difference $\omega-\omega'$.}
 \label{fig:velocity}
\end{figure}

Fig.~\ref{fig:velocity} shows the measured mean velocity of atoms along the $z$ axis as a function of
the frequency difference $\omega-\omega'$ between the two pairs of trapping beams. As expected from
(\ref{eq1}), we find a linear relation with a slope in agreement with the imposed geometry
($\alpha=47^\circ$).

\begin{figure}
 \includegraphics[width=0.45\textwidth]{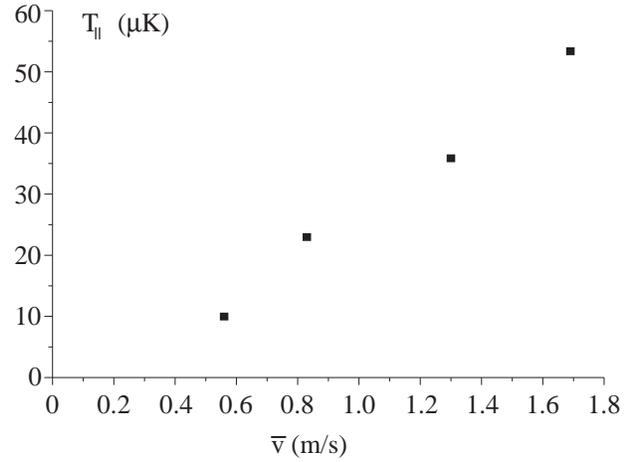}
\caption{Longitudinal temperature $T_\parallel$ as a function of
the mean velocity $\bar v$ of the beam.}
 \label{fig:temperature}
\end{figure}

We have measured the longitudinal temperature $T_\parallel$ as a
function of the mean velocity. The results are shown in
Fig.~\ref{fig:temperature}. The longitudinal temperature increases
when the mean velocity increases. The atomic beam which is
produced in this experiment is quasi mono-kinetic. We find that
$\Delta v_{\parallel}/\bar{v} < 0.1$ over the whole range
0.3--3~m/s for $\bar v$. The lowest temperature
$T_\parallel=10\;\mu$K is obtained for $\bar v=60$~cm/s.

This temperature is significantly smaller than what is expected
from standard laser cooling theory, for our trapping beam
intensity and detuning. Such a low temperature probably results
from an additional cooling of the atoms as they slowly move out
from the trapping beams. A similar effect occurs for 3D-MOT in the
time domain, when one switches relatively slowly the MOT beams
(see \cite{Weiss90,Salomon91}). In the context of slow atomic
beams generated from a MOT, it has also been observed in
\cite{Chen00}.

\subsubsection{Transverse temperature of the beam}

The geometry of the MM-MOT creates a strong asymmetry between the longitudinal $z$ axis and the
transverse $xy$ axes. For $\alpha\sim 0$, the cooling is much more efficient along the $z$ axis than
in the $xy$ plane, and the conclusion is reversed when $\alpha\sim\pi/2$. Therefore there is no
reason to expect that the two velocity distributions are characterized by the same temperature,
unless the elastic collision rate is large enough to thermalize all degrees of freedom of the atomic
beam. However, the spatial densities achieved so far in our experiment are too low for this
thermalization to occur when the atoms travel over the distance $z_0$.

We have measured directly the transverse size of the atomic beam
by scanning the probe beam position. In this way, we deduce an
upper bound for the transverse temperature. We find $T_\perp=35
\;\mu$K for $\bar{v}=0.3$ m/s and $T_\perp=80 \;\mu$K for
$\bar{v}=1$ m/s. Again, this variation of the temperature with the
longitudinal velocity can be attributed to the extra cooling which
occurs when the atoms leave the trapping region slowly enough.

\subsubsection{Loading and steady-state flux of the atomic source}

The loading dynamics of the MM-MOT can be understood with a simple model based on the fact that even
for the highest average velocities $\bar v$ we have produced ($\bar v$=3~m/s), the average time
$L/(2\bar v)$ required for an atom to leave the MM-MOT ($L \sim 2$~cm is the length of the atomic
cloud) is larger than the transverse trapping time deduced from our simulation ($\sim 2$~ms).
Therefore we can assume that the loading rate $R$ of the MM-MOT (number of atoms captured per unit
length and unit time) is independent of the velocity $\bar v$, for the velocity range considered
here. Consider now a point inside the MM-MOT close to the trap exit. The time dependence of the
linear density in this point is expected to vary as
\begin{equation}
n(t) = R \int_0^{\min(t,L/\bar v)}e^{-\gamma t'}\;dt'\ ,
\end{equation}
where $\gamma$ denotes the loss rate of atoms from the trap, \emph{e.g.} due to collisions with the
residual gas. In particular, the steady state value of this quantity is
$n(\infty)=R/\gamma(1-\exp(-\gamma L/\bar v)) \simeq R L/\bar v$ if the lifetime $\gamma^{-1}$ is
much larger than the ``launching time" $L/\bar v$. This corresponds to a steady-state atom flux
$\Phi=RL$ independent of the mean velocity $\bar v$.

\begin{figure}
\includegraphics[width=0.45\textwidth]{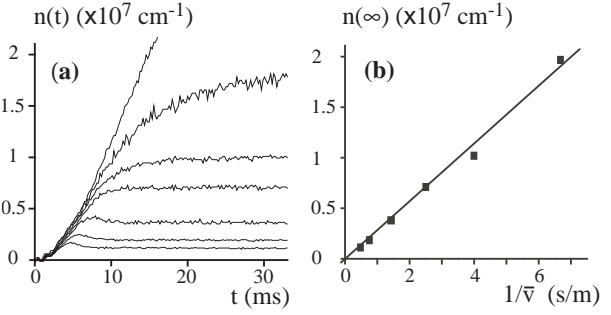}
 \caption{(a) Time variation of the linear density $n(t)$ at
the exit of the MM-MOT, inferred from the fluorescence intensity at this location, for various
launching velocities $\bar v$ (from top to bottom, $\bar v=$ 0, 15, 25, 40, 70, 130, 210 cm/s). (b)
Asymptotic linear density $n(\infty)$ as a function of the launching velocity $\bar v$.}
 \label{fig:loading}
\end{figure}

The measurement of the linear density of trapped atoms confirms
this simple model, as can be seen in Fig.~\ref{fig:loading}. This
measurement is performed by analyzing the fluorescence signal
emitted from a small region located close to the trap exit. In
order to avoid difficulties related with a residual magnetic field
or with inhomogeneities in the laser beam profiles, we calibrate
the fluorescence measurement by suddenly bringing the trapping
laser frequency in resonance with the atoms, so that each atom
emits $\sim \Gamma/2$ photons/s for a short time before being
expelled.

The linear increase of the signal at short times shown in
Fig.~\ref{fig:loading}a is the same for all velocity classes and
it gives $R\simeq  10^9$~atoms/s/cm for a $^{87}$Rb rubidium vapor
pressure $P_{87}=3\;10^{-9}$~mbar. For long loading times, the
linear density $n(t)$ saturates to a value $n(\infty)$ which is
approximately proportional to $1/\bar v$
(Fig.~\ref{fig:loading}b). This shows that the steady state flux
$\Phi=\bar v \;n(\infty)$ is independent of the average velocity
of the beam, as predicted by the simple model described above.
% ($\Phi=3\;10^{8}$~atoms/s).
This flux is
proportional to $P_{87}$ in the range $10^{-9}$~mbar $< P_{87} < 10^{-8}$~mbar, and we obtain:
\[
\Phi= 10^9 \mbox{ atoms/s for } P_{87}=10^{-8} \mbox{ mbar.}
\]

The measured flux is in good agreement with the predictions of the simulation described in \S
\ref{subsec:PrincipleMMMOT}. For the same laser intensities and waists, the simulation gives after
averaging over the atomic trajectories a capture velocity $v_{\rm cap}=34$~m/s, corresponding to a
flux $\Phi=2.2\; 10^9$~atoms/s for $P_{87}=10^{-8}$~mbar. Since $\Phi \propto v_{\rm cap}^4$, this
means that the simulation overestimates the capture velocity by 20\%. This slight probably results
from the approximations which led to the expression (\ref{Fsum}) for the radiative force in the
MM-MOT.

\subsection{Comparison with previous works}

As already mentioned in the introduction, our experimental setup for the MM-MOT  is quite different
from the devices of the leaking MOT family. These devices all lead to an average velocity which is at
least one order of magnitude larger than the lowest velocity demonstrated here, and they could not be
used as a convenient source for the magnetic guide in which we want to perform evaporative cooling.
Our source presents certain similarities with the ones discussed in \cite{Swanson96,Chen00,Weyers97}.
In these experiments, a magnetic configuration similar to ours was used and a moving molasses was
produced along the beam axis, while two-dimensional magneto-optical trapping confined the atoms
transversally. However, some important differences should also be pointed out between these
experiments and our setup.

(i) The two experiments \cite{Swanson96,Chen00} dealt with a decelerated atomic beam while our MM-MOT
is directly loaded from the residual rubidium vapor pressure in our vacuum cell. This is an important
simplification for our setup, yet there are some drawbacks. Since our purpose is to load a magnetic
guide, we have to maintain a vacuum which is good enough to prevent collisional loss of atoms while
they travel from the MM-MOT to the entrance of the magnetic guide. Typically we have to restrict the
$^{87}$Rb pressure to a value below $10^{-8}$~mbar, for an atom transit time of the order of 0.1~s.
For this background pressure, our flux is respectively 6 times smaller and 2 times larger than what
is reported in \cite{Swanson96} and \cite{Chen00}. The experiment described in \cite{Weyers97} was
based on a vapor loaded trap as ours, but the reported flux was three orders of magnitude lower than
the present one, which makes the design of \cite{Weyers97} quite impracticable for our purpose.

(ii) We use a laser configuration with only four beams, instead of six. This is known as a very
robust configuration for laser cooling and trapping processes. Any phase fluctuation of the laser
beams only results in a translation of the whole interference pattern of the trapping beams
\cite{Grynberg93}, whereas in a five-beam or a six-beam configuration, this pattern changes when one
of the phases of the beams fluctuates giving rise to heating. So far, we do not benefit from this
increased stability, but it may play a role in future work. (iii) Finally we note that we have been
able to operate our MM-MOT with exit velocities $\bar v$ notably smaller than what is reported in
\cite{Swanson96,Chen00,Weyers97} where $\bar v \geq 2$~m/s. This is important for the future
implementation of forced evaporative cooling along the magnetic guide into which the atomic beam is
injected. For a given length of the guide, a smaller velocity leads to an increased evaporation time,
and therefore a better cooling.

\section{Loading of the magnetic guide}
\label{sec:guide}

As explained in \S~\ref{sec:introduction}, the second important result of the present paper concerns
the injection of the slow atomic beam described above into a magnetic guide. This is an essential
step towards the long term goal of this work, which consists in cooling the guided atoms by
radio-frequency evaporation.

\subsection{Characteristics of the guide}

\begin{figure}
 \includegraphics[width=0.45\textwidth]{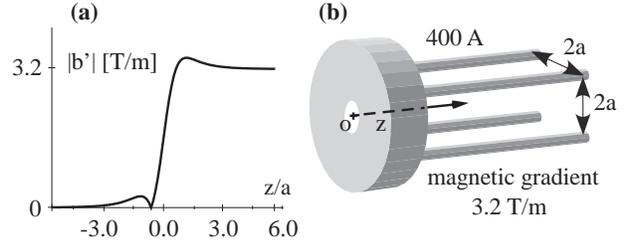}
 \caption{Entrance of the magnetic guide. (a) Variation of the magnetic gradient
in the $xy$ plane as a function of $z$. (b) Schematic drawing of the guide. The cylindrical part
located in close to $z=0$ allows for the connection of the electrical currents and the water cooling
circuit circulating into the four copper tubes.}
 \label{fig:guide}
\end{figure}

The magnetic guide has a total length of 60~cm. It is made out of
four copper tubes (\O$_{\rm ext}=6$~mm and \O$_{\rm int}=4$~mm)
placed in quadru\-pole configuration (see Fig. \ref{fig:guide}a).
The tubes are located in the domain $z>0$ and they are joined in
$z=0$ by a hollow metal cylinder, which allows for the circulation
of current and cooling water from tube to tube. The axes of the
copper tubes are placed at coordinates $x=\pm\, a; y=\pm\, a$,
with $a=7\;$~mm. A current $I=400$~A can be sent through the
tubes, which provides, far from the entrance of the guide
(\emph{i.e.} $z \gg a$), a magnetic gradient $b'=3.2$~T/m in the
$xy$ plane.

The value of $b'$ as a function of $z$ is plotted on
Fig.~\ref{fig:guide}b. It changes sign in $z=-3$~mm and its
absolute value is reduced by a factor larger than 100 for $z=
-25$~mm, where the exit of the MM-MOT is located. Therefore this
magnetic gradient does no affect the operation of the MM-MOT.

It is possible to superimpose a bias field parallel with the $z$ axis to prevent Majorana spin flips
when a guided atom passes close to the axis of the trap $x=y=0$, where the trapping field is zero. In
practice we have found that this bias field does not play a significant role, which can be easily
understood by the following estimate. The probability for a spin flip when an atom passes close to
the trap axis is $p\sim 2\,e^{-\xi}$, where $\xi=\pi \mu_B b' \rho^2/(4\hbar v_\bot)$
\cite{Landau,Vitanov97}. Here $\rho$ is the minimal distance from the $z$ axis (\emph{i.e.} the
impact parameter), $v_\bot$ designates the velocity in the $xy$ plane at this location, and we assume
that the atoms are trapped in the hyperfine state $F=1,m=-1$, whose magnetic moment $\mu$ is half the
Bohr magneton $\mu_B$. Taking as a typical value $v_\bot\sim 0.1$~m/s, we find that $p$ is important
only if $\rho \leq 1\;\mu$m. Therefore this loss term is not expected to play a role in our
experiment, since the average impact parameter is $\geq 100\;\mu$m and the elastic collision rate is
negligible. On the contrary, for higher atomic densities, elastic collisions can reload the phase
space cells corresponding to trajectories with very small impact parameters. These losses should then
play an important role, especially in our 2D quadrupole field where they occur close to a line, by
contrast with a 3D quadrupole trap where they occur only around the central point. The presence of a
bias field should then be crucial.

A narrow glass tube (\O 4 mm, length 170 mm) is centered at the entrance of the magnetic guide and
aligned with the $z$ axis. It ensures differential pumping between the MM-MOT cell, where the Rb
partial pressure must be large enough to provide an efficient loading, and the chamber of the
magnetic guide, where the residual pressure has to be minimized to avoid losses due to collisions
with the background gas. We estimate that the pressure in this chamber is $\sim 4\;10^{-10}$~mbar,
which corresponds to a lifetime $\sim 20$~s for the magnetically guided atoms.

\subsection{Connection of the MM-MOT to the guide in continuous mode}
\label{subsec:connectioncontinuous}

The simplest way to load atoms from the MM-MOT into the guide
consists in operating the MM-MOT continuously at a fixed velocity
$\bar v$. As explained in \S~\ref{sec:MMMOT}, the angle of the
MM-MOT beams with the $z$ axis is $\sim \pi/4$. Due to the
presence of the metal cylinder located at the entrance of the
magnetic guide (see Fig.~\ref{fig:guide}a), the shortest distance
between the output of the MM-MOT and the entrance of the magnetic
guide is $D\sim 25$~mm. In other words, the MM-MOT is located in
the region $-45$~mm $<z<-25$~mm and the atoms have to travel over
a distance of $25$~mm before being captured by the magnetic guide
if they are in a magnetic sublevel corresponding to a low-field
seeking state.

\subsubsection{Free flight transfer}

The simplest connection mode relies on ballistic flight to ensure the transfer between the MM-MOT and
the magnetic guide. We place a laser tuned to the $|5S_{1/2},F=2\rangle\rightarrow
|5P_{3/2},F=2\rangle$ transition in the region $-25$~mm~$<z<0$. The repumping laser is blocked in the
transfer region, so that this extra ``depumping" laser beam optically pumps the atoms in the $F=1$
ground state. Therefore, provided that the three magnetic levels $m=0,\pm 1$ of the
$|5S_{1/2},F=1\rangle$ are equally populated, we expect that one third of the atoms (state $m=-1$)
emerging from the MM-MOT can be captured by the magnetic guide. This method based on ballistic flight
is in principle quite simple, but has several drawbacks:
\begin{itemize}
\item
It is restricted to relatively large velocities $\bar v >
1.5$~m/s. Otherwise the free-fall due to gravity, $gD^2/(2\bar
v^2)$, is so large that the coupling to the guide becomes
velocity-dependent .

\item
During the free flight, the atom jet spreads transversally, which leads to a strong increase of the
transverse temperature in the guide, as compared with the MM-MOT. Consider for instance a beam
emerging from the MM-MOT with $\bar v=2$~m/s and $T_\bot=100\;\mu$K. At the entrance of the magnetic
guide, the radius of the beam is $R =(D/\bar v)\,\sqrt{k_BT/M}\sim 1.2$~mm ($M$ is the atomic mass).
The corresponding magnetic energy is
\begin{equation}
E_{\rm mag}=\mu b'R=(D\mu b'/\bar v)\,\sqrt{k_B T/M}\ .
\label{temperature}
\end{equation}
This gives  $E_{\rm mag}/k_B\sim 1$~mK in the present case.

\item
The optical pumping of the atoms to the hyperfine level $F=1$ is not total, because of the stray
light present at the repumping frequency $|5S_{1/2},F=1\rangle\rightarrow |5P_{3/2},F=2\rangle$.
Therefore, after leaving the MM-MOT, the atoms may still be deflected by the residual radiation
pressure force resulting from an imbalance between the intensities of the various laser beams.
Although we took a particular care for adjusting the spatial profiles of these laser beams along the
path corresponding to the trap exit, we found that it is very difficult to transfer slow atoms ($\bar
v < 2$~m/s) with this method.
\end{itemize}

\subsubsection{Use of an auxiliary transfer MOT}

\begin{figure}
\includegraphics[width=0.45\textwidth]{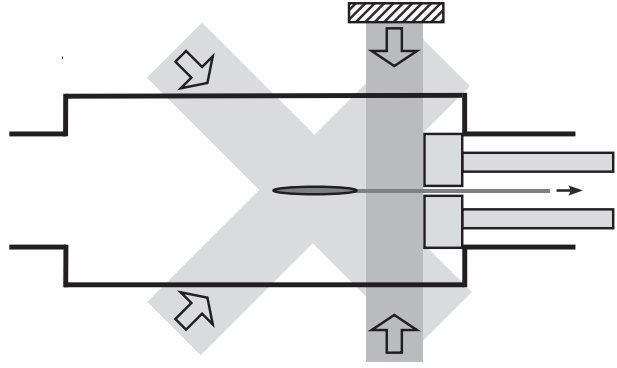}
\caption{Use of an auxiliary 2D MOT to
transfer efficiently the atoms from the MM-MOT to the magnetic guide.}
 \label{fig:transferMOT}
\end{figure}

The difficulties mentioned above can be circumvented if we place
an auxiliary guiding trap in the region $-25$~mm $<z<0$. This trap
is a pure 2D MOT, whose beams which are orthogonal to the $z$ axis
illuminate points up to 4~mm to the entrance of the magnetic guide
(see Fig.~\ref{fig:transferMOT}). The transverse extension of the
atomic beam at the entrance of the magnetic guide is then
drastically reduced. As for the free flight transfer, we place a
depumping laser at the entrance of the guide. The atoms are then
trapped in the magnetic sublevel $J=1,m=-1$, for which the amount
of resonant stray light is too low to be an appreciable source of
losses in the guide .

The disadvantage of this auxiliary trap concerns the broadening of the longitudinal velocity
distribution, due to the random recoils associated with the spontaneous emission of photons as the
atoms cross this auxiliary trap. However, as we show below (\S~\ref{subsec:positionguide}), this
longitudinal heating remains relatively weak for a large detuning $\delta_2$ of the auxiliary MOT. An
upper limit for $|\delta_2|$ is given only by the condition that the radiative force of the auxiliary
MOT has to overcome the spurious radiation pressure imbalance which exists in the wings of the MM-MOT
lasers, as discussed above.

\subsection{Connection of the MM-MOT to the guide in pulsed mode}

The operation of the MM-MOT in a continuous mode, as described in
the previous section (\S~\ref{subsec:connectioncontinuous}),
imposes to search for a compromise between two different
requirements. For an efficient loading one has to take a
relatively small detuning ($\delta \sim -3\,\Gamma$), so that the
trapping force is large. On the contrary, in order to minimize the
temperature of the outgoing atomic beam, a much larger detuning
$(\delta \sim-7\,\Gamma$ for our laser intensity) is preferable.

A pulsed operation of the MM-MOT may provide, for a given application, an output beam with better
characteristics than this compromise. First, one loads the MM-MOT with the detuning which maximizes
the capture rate $R$ and with a zero launch velocity ($\omega=\omega'$). During this phase of
duration $t_1$, the number of trapped atoms varies according to $N(t)=(RL/\gamma)\,(1-e^{-\gamma
t})$. One then switches the detuning of the trapping lasers to a much larger value, with $\omega \neq
\omega'$ set to provide the required velocity $\bar v$. This launching phase must last a time
$t_{2}=(L+D)/\bar v$, so that all trapped atoms can leave the MM-MOT and reach the magnetic guide.
The optimization of $t_1$ depends on the relative value of the escape time $\gamma^{-1}$ and the
launching time $t_2$. For a low Rb vapor pressure (small $R$ and small $\gamma$), the largest flux
corresponds to
\begin{equation}
\gamma t_2 \ll 1 \quad : \qquad \Phi \simeq RL \qquad \mbox{for}
\qquad t_1=\sqrt{2t_2/\gamma}\ . \label{optimumflux}
\end{equation}
In this case, the flux $\Phi$ in the guide is equal to the capture rate of the MM-MOT and the
operation in pulsed mode does not lead to a loss in efficiency. If we increase the Rb vapor pressure
in the cell and thus the rate $R$ so that $\gamma \sim t_2^{-1}$, the optimum operation of the pulsed
mode corresponds to
\begin{equation}
\gamma t_2 \sim 1 \quad : \qquad \Phi \sim 0.3\,R L \quad
\mbox{for} \quad t_1 \sim t_2 \sim \gamma^{-1} \ .
\label{optimumflux2}
\end{equation}

This pulsed mode makes the extraction of atoms from the MM-MOT much easier. In the continuous mode,
the detuning of the MM-MOT laser beams is relatively small ($\delta \sim -3 \Gamma$) so that the
outgoing atoms may be deflected or accelerated by the fringe field of the trapping laser beams, as
already described. This spurious effect is strongly decreased if the detuning of these lasers is
increased to a value $\delta \sim -7\,\Gamma$ while the atoms travel in the ``dangerous" region.

We note that, as for the continuous mode of operation, the loading of the magnetic guide from the
MM-MOT may be improved by an auxiliary 2D MOT located between the exit of the MM-MOT and the guide.
Also, although the atomic beam is now pulsed at the entrance of the guide, the pulses broaden as they
propagate because of the longitudinal velocity dispersion $\Delta v$. This entails that a
quasi-continuous beam is obtained after a distance $\sim L \bar v/\Delta v$ if one chooses $t_1\sim
t_2$.

\subsection{Position and velocity distribution of the guided atoms}
\label{subsec:positionguide}

We have studied experimentally the various ways for loading the
magnetic guide discussed above, \emph{i.e.} without or with the
auxiliary 2D MOT, and for a continuous or pulsed operation of the
MM-MOT. As for the MM-MOT itself, we have determined the
longitudinal and transverse temperatures of the atoms in the
guide, as well as the flux of the guided beam.

\subsubsection{Longitudinal velocity distribution}

\begin{figure}
\includegraphics[width=0.45\textwidth]{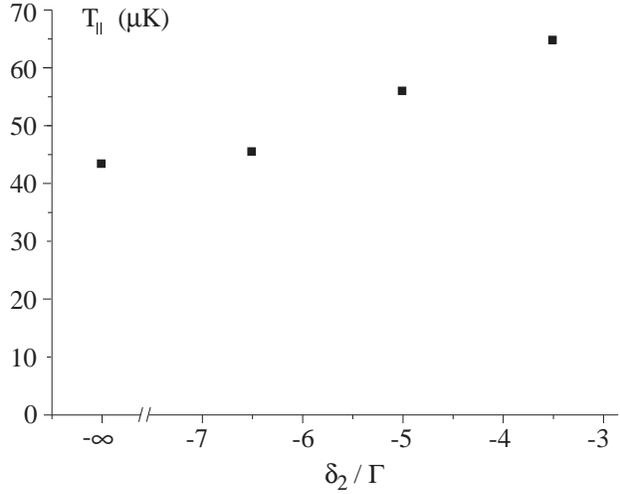}
 \caption{Longitudinal temperature in the
magnetic guide, as a function of the detuning $\delta_2$ of the auxiliary MOT. The value found in
absence of the auxiliary MOT is given on the left of the graph. For these measurements, the MM-MOT is
operated in pulsed mode with $\bar v=2.6$~m/s, $b'=1.5$~T/m. }
 \label{fig:longitudinalguide}
\end{figure}

The measurement of the longitudinal velocity distribution is quite
straightforward if we operate the MM-MOT in pulsed mode, since we
can derive the longitudinal temperature directly from the temporal
width of the absorption signal when a given atom pulse passes
through the probe beam. This beam is located downstream, at 40~cm
from the entrance of the magnetic guide. The results are given in
Fig.~\ref{fig:longitudinalguide}, as a function of the detuning
$\delta_2$ of the auxiliary MOT. This figure also gives the
results obtained in the absence of the auxiliary MOT.

We find that the temperatures are always lower for larger velocities ($\bar v=2.6$~m/s). This is the
opposite of what we found just at the output of the MM-MOT. It is probably due to the heating of the
atoms by the stray light of the various trapping beams, when they travel over the distance $D$
between the MM-MOT and the guide. This heating is larger if the atoms spend more time in this region,
\emph{i.e.} if they are slow. The presence of the auxiliary MOT causes some additional heating of the
atomic beam which is also proportional to the time spent by the atoms in the laser beams. For
$\delta_2=-6.5\;\Gamma$, the increase of the longitudinal temperature of the beam is at most
$30\,$\%, while, as we shall see, the transverse temperature decreases by an order of magnitude,
thanks to the auxiliary MOT.

When the MM-MOT is operated in continuous mode, we found much larger longitudinal temperatures, in
the range of 0.5--1~mK. We think that this is due to the acceleration and heating of the atoms when
they travel between the MM-MOT and the guide, the heating being much more dramatic than in the pulsed
mode, since the trapping beams are much closer to resonance.

\subsubsection{Transverse temperature}

The transverse temperature in the guide has been measured in two
ways, which give consistent results. First, we can use the method
already described for the characterization of the MM-MOT which
consists in scanning the position of the probe laser beam, in
order to reconstruct the transverse profile of the atomic beam.
Alternatively, we can use radio-frequency evaporation to
selectively remove a fraction of the atomic distribution. From the
variation of the fraction of remaining atoms as a function of the
radio-frequency, we can infer the transverse temperature of the
atomic beam.

\begin{figure}
\includegraphics[width=0.45\textwidth]{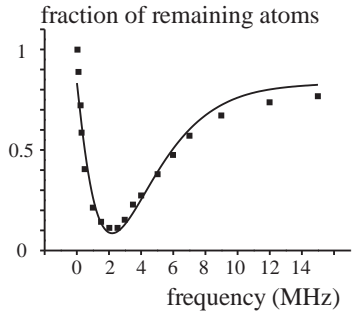}
\caption{Fraction of remaining atoms as a
function of the radio-frequency applied in a region located between the entrance of the guide and the
detection region. From this set of results, we deduce the transverse temperature of the guided atomic
beam, here $170\,\mu$K. Here the MM-MOT is operated in pulsed mode with $\bar v=2.6$~m/s,
$b'=1.5$~T/m. The detuning of the auxiliary MOT is $\delta_2=-6.5\;\Gamma$. }
 \label{fig:transverseguide}
\end{figure}

A result obtained with this second method is shown in
Fig.~\ref{fig:transverseguide}. It has been obtained with a pulsed
loading in presence of the auxiliary MOT. We find that the losses
of atoms are maximal for a radio-frequency $\sim\;2$~MHz. Using a
simple numerical model based on a Monte-Carlo sampling of the
atomic distribution, we can determine from this set of data the
transverse temperature of the beam, which is in this case $\sim
170\;\mu$K.

We have measured the transverse temperature of the magnetically guided atoms in the presence of the
auxiliary MOT, using the MM-MOT either in the continuous or the pulsed regime and we have obtained
similar results. The variation of the transverse temperature with the magnetic gradient $b'$ and the
longitudinal velocity is in good agreement with the prediction (\ref{temperature}), where $D$ should
be replaced by the distance $D'\sim 4$~mm between the exit of the auxiliary MOT and the entrance of
the magnetic guide. We did not find a strong dependence of the transverse temperature of the guided
atoms on the detuning $\delta_2$ of the auxiliary 2D-MOT, which can be understood easily. When
$|\delta_2|>5\;\Gamma$, one expects transverse temperatures within the 2D MOT in the $10\,\mu$K
range. The corresponding kinetic energy is negligible compared with the energy (\ref{temperature})
resulting from the divergence of the atomic beam, as the atoms travel over the distance $D'$.

In absence of the auxiliary MOT, we have measured much larger transverse temperatures inside the
guide, in the range of 1-2~mK. This is a consequence of the increase of the beam's transverse size at
the magnetic guide entrance, after free propagation over the distance $D$. The variation of the
transverse temperature with $b'$ and $\bar v$ is also in good agreement with the prediction
(\ref{temperature}).

\subsubsection{Flux}

The flux of atoms in the magnetic guide is not significantly modified by the presence of the
auxiliary MOT. Operating in continuous mode at a pressure $P_{87}=10^{-8}$~mbar, we find that this
flux varies between $1.5\;10^8$ and $3\;10^8$~atoms/s, when $\bar v$ varies between $1.5$~m/s and
$3$~m/s. This corresponds to a transfer efficiency between 15\% and 30\%. Similar values are achieved
in pulsed mode if we optimize the loading time $t_1$ according to (\ref{optimumflux}).

For velocities smaller than 1.5~m/s, we did not find that a significant fraction of the atoms emitted
by the MM-MOT could be transferred to the magnetic guide. We attribute this fact to the depletion of
the slow atomic beam by collisions with atoms from the rubidium vapor in the cell. The slow atoms
have to travel over a distance $\sim 10$~cm through the cell and the tube ensuring the differential
pumping, before they arrive in the ultra-high vacuum region of the magnetic guide. For
$P_{87}=10^{-8}$~mbar (i.e. a total rubidium pressure $4\;10^{-8}$~mBar) and $\bar v=1.5$~m/s, we
estimate the atomic flux to be reduced by 40~\% over this distance. This decay factor is expected to
vary as $\exp(-\alpha P_{87}/\bar v)$ (where $\alpha$ is a numerical constant), which we check
experimentally by varying $P_{87}$ between $10^{-9}$~mbar and $10^{-8}$~mbar and $\bar v$ between
1~m/s and 3~m/s. The decay makes the guided beam very difficult to detect for ultra-low velocities
($\bar v \leq 1$~m/s), especially if one also takes into account the increase of the transverse
temperature for low $\bar v$ (see eq.~(\ref{temperature})). On the other hand, for relatively large
velocities ($\bar v\geq 2.5$~m/s), the flux of guided atoms is close to the one expected from a
uniform repartition of the atoms among the three possible Zeeman sublevels (the untrapped levels
$m=1,0$ and the trapped one $m=-1$), \emph{i.e.}, 1/3 of the atoms emerging from the MM-MOT.

\section{Conclusion and perspectives}
\label{sec:conclusion}

We have presented in this paper the experimental study of the
production of a slow atomic beam from a ``moving molasses $+$
magneto-optical trap" setup (MM-MOT), and the transfer of this
beam into a magnetic guide. This setup can be operated in
continuous as well as in pulsed mode. In particular, our results
prove that the light scattering from the atoms being captured in
the MM-MOT can be kept low enough to avoid perturbations of the
propagation of the atoms in the guide. This is a crucial point for
future work towards the goal of this project, which consists in
performing evaporative cooling in the magnetic guide. Our approach
constitutes an alternative to the multiple loading of a 3D
magnetic trap \cite{Cornell91,Davies00}.

Due to the particular geometry of our setup, the best transfer
between the MM-MOT and the magnetic guide is ensured by using an
auxiliary transverse 2D-MOT which nearly fills the $D=25$~mm gap
between the end of the MM-MOT and the entrance of the magnetic
guide. The price to pay for the gain in transverse confinement is
a slight increase of the longitudinal temperature of the atoms,
due to the recoil heating along the beam axis. This auxiliary
2D-MOT will be unnecessary in the new version of our setup where
the distance $D$ will be considerably reduced.

The lowest longitudinal temperature of the guided atoms has been
obtained by operating the MM-MOT in pulsed mode, \emph{i.e.} by
switching the detuning of the MM-MOT laser beams to a large value
(typically seven line widths) during the launching phase. In this
way, we can minimize the acceleration of the atoms by the laser
stray light as they travel over the distance $D$. For a
longitudinal velocity $\bar v=2$~m/s, the measured longitudinal
temperature is $\sim 50\;\mu$K. The transverse temperature is
significantly larger ($\sim 170\;\mu$K), as a consequence of the
divergence of the atomic beam between the exit of the MOT region
and the entrance of the magnetic guide. The transverse spatial
spreading of the atomic beam translates into an increase of the
transverse temperature, as expressed by (\ref{temperature}).

The largest atom flux in the guide is $\sim 3\;10^8$~atoms/s, and it is limited by the loading flux
of the MM-MOT, which is itself proportional to the $^{87}$Rb vapor pressure in the cell. Going to
higher pressures would improve the loading flux, but it would also increase the collisional losses of
the slow atomic beam on its way to the magnetic guide. A decisive improvement in this context is to
turn to a MM-MOT loaded from an auxiliary slow atomic beam. This increases the complexity of the
setup, but should allow for a ten times larger guided flux.

The continuous operation of our setup is in contrast with the pulsed loading of a magnetic guide as
reported in \cite{Teo01}. In that work, $10^5$ atoms were coupled into a four-wire guide from a
3D-MOT, whereas the pulsed operation of our present system gives up to $10^7$ atoms/cycle, with
comparable longitudinal and transverse temperatures.

Finally, we note that the collision rate on axis in our setup, which scales as $\Phi/(\bar v T)$, is
at present $\sim 3\;10^{-2}$~s$^{-1}$. This is still much too small to initiate evaporative cooling.
However, we expect to increase the flux by a factor 10 using an auxiliary slow atomic beam. Another
factor 10 should be gained from an improved transfer scheme from the MM-MOT into the magnetic guide,
since this will decrease the temperature $T$ of the trapped atoms and allow to work at a lower
average velocity $\bar v$. In these conditions, an adiabatic compression of the atomic beam within
the guide, using for instance permanent magnets, should lead to a collision rate in the usual range
for compressed magnetic traps, compatible with a spatial evaporation along a guide of reasonable
length.

\begin{acknowledgement} C. F. Roos acknowledges support from the European Union (contract HPMFCT-2000-00478). A. Aclan was supported by a grant from Volkswagen and the Gottlieb Daimler and
Karl Benz-Stiftung. This work was partially supported by CNRS, the R\'egion Ile-de-France, Coll\`{e}ge de France and DRED.
\end{acknowledgement}

\end{document}